\documentclass[pre,twocolumn,showpacs,floatfix]{revtex4}
\usepackage{graphicx}
\usepackage{amsmath}
\usepackage{amssymb}
\usepackage{subfigure}
\newcommand{\ww}[1]{\underline{\underline{{\bf #1}}}}

\newcommand{\be}{\begin{equation}}
\newcommand{\ee}{\end{equation}}
\newcommand{\bea}{\begin{eqnarray}}
\newcommand{\eea}{\end{eqnarray}}
\newcommand{\ba}{\begin{aligned}}
\newcommand{\ea}{\end{aligned}}
\newcommand{\bma}{\begin{bmatrix}}
\newcommand{\ema}{\end{bmatrix}}

\newcommand{\norm}[1]{\vert \vert {\bf #1}\vert \vert}
\newcommand{\normm}[1]{\vert \vert #1\vert \vert}
\newcommand{\bi}{\begin{itemize}}
\newcommand{\ei}{\end{itemize}}
\newcommand{\im}{\item}

\newcommand{\tr}{\text{tr}\hspace{.05em}}
\newcommand{\disty}{\displaystyle}
\newcommand{\ave}[1]{\langle #1 \rangle}
\begin{document}

\title{Internal states of model isotropic granular packings.\\
II.  Compression and pressure cycles.}

\author{Ivana Agnolin}
\author{Jean-No\"el Roux}
\email{jean-noel.roux@lcpc.fr}
\affiliation{Laboratoire des Mat\'eriaux et des Structures du G\'enie Civil\footnote{
LMSGC is a joint laboratory depending on Laboratoire Central des Ponts et Chauss\'ees, \'Ecole Nationale
des Ponts et Chauss\'ees and Centre National de la Recherche Scientifique},
Institut Navier, 2 all\'ee Kepler, Cit\'e Descartes, 77420 Champs-sur-Marne, France}

\date{\today}

\begin{abstract}
This is the second paper of a series of three investigating, by numerical means, the geometric and mechanical properties 
of spherical bead packings under isotropic stresses. We study the effects of varying the applied pressure $P$ (from 1 or 10~kPa up to 100~MPa in the
case of glass beads) on 
several types of configurations assembled by different procedures, as reported in the preceding paper~\cite{iviso1}. 
As functions of $P$, we monitor changes in solid fraction $\Phi$, coordination number $z$, proportion of \emph{rattlers}
(grains carrying no force) $x_0$, the distribution of normal forces, the level of friction mobilization, and the distribution of
near neighbor distances. Assuming that the contact law does not involve material plasticity or damage, $\Phi$ is found to vary very nearly reversibly with $P$
in an isotropic compression cycle, but all other quantities, due to the frictional hysteresis of contact forces, change irreversibly. In particular, initial
low $P$ states with high coordination numbers lose many contacts in a compression cycle, and end up with values of $z$ and $x_0$ close to those of the most
poorly coordinated initial configurations. 
Proportional load variations which do not entail notable configuration changes can therefore nevertheless
significantly affect contact networks of granular packings in quasistatic conditions.

\end{abstract}


\maketitle

\section{Introduction}
The mechanical properties of solidlike granular packings are traditionally studied, at the macroscopic level, in
engineering fields such as soil mechanics~\cite{DMWood,BiHi93,MIT93,HHL98}, and are currently being investigated, with some attention to the 
grain scale and micromechanical origins of macroscopic behaviors, in condensed matter physics and material science communities~\cite{HHL98,HW04,GRMH05}. 

The present paper, the second of a series of three, investigates, by numerical simulations, 
the mechanical and microstructural response of a model material, the packing of identical spherical
beads, to pressure intensity variations. It refers a lot to the results of the previous, companion paper~\cite{iviso1}, but may be read independently.

Although molecular dynamics (or ``discrete element'') approaches have repeatedly been applied to sphere
packings~\cite{MGJS99,TH00,MJS00,SEGHL02,SGL02,SUFL04}, many important questions related to the microscopic origins
of their macroscopic mechanical behavior in the quasistatic regime have not been fully explored yet. One such issue is the influence of the initial state,
which is determined by the assembling process. 
In the first paper of the present series~\cite{iviso1} (hereafter referred to as paper I),
the results of several packing preparation methods, all producing ideally isotropic states, are compared. Direct compressions of granular
gases produce states that do not depend on dynamical parameters if the compression is slow enough. Their solid fraction $\Phi$ and coordination number $z^*$
(evaluated on excluding the rattlers, a proportion $x_0$ of grains which do not carry any force) are decreasing functions of the friction coefficient $\mu$, from
$\Phi\simeq 0.639$ and $z^*=6$ for $\mu=0$, in which case the random close packing state (RCP) 
is obtained, down to $\Phi\simeq 0.593$ and $z^*\simeq 4.5$ for $\mu=0.3$.
In paper I~\cite{iviso1} we accurately checked the uniqueness of the RCP, on confronting our own numerical results with those of
several recent publications, in which different numerical procedures were implemented~\cite{DTS05,OSLN03}. In the presence of intergranular friction, however,
quite different packing states might be prepared.
First, it is of course possible, in a simulation, to increase the friction coefficient once the packing is equilibrated under some pressure; such
a numerical procedure can be regarded as a model for an assembling process in the presence of a lubricant within intergranular gaps in the laboratory. Ideally, 
whatever the value of the friction coefficient used to model the quasistatic mechanical properties of the material, 
it is possible to assemble the sample 
with $\mu=0$ (thus assuming ideal, perfect lubrication in the fabrication stage) and hence with the RCP density and coordination number. Once the grains are packed
and form a solid material, contacts between grains can then be attributed the final, finite friction coefficient used in quasistatic modelling.
Experimentally, it is of course well known that given granular materials can be packed with varying densities. 
A common method to make them denser, other then lubricating the contacts in the assembling stage, is 
the application of vibrations or ``taps''. A numerical idealized vibration procedure, apt to prepare dense samples
with little computation time, was defined in paper I. Surprisingly, although it produces isotropic states with densities close to the RCP value, 
their coordination numbers are
as low as in the loosest states assembled by direct compression. The small geometric differences between configurations with the same solid fraction but very 
different coordination numbers is still not accessible to tomographic observation techniques~\cite{iviso1}. 
Only mechanical properties can thus be confronted to experimental results, 
to determine whether or in which conditions the investigated numerical systems are close to experimental reality. 

Before studying elastic properties in paper III of the present series~\cite{iviso3}, one should first investigate the effect of an isotropic compression.
The application of a large enough confining pressure, usually at least a few tens of kPa (with rare 
exceptions~\cite{Tat39,ReCl01}), is necessary before the macroscopic mechanical behavior of solidlike granular packings
is tested~\cite{VER98,GDV84,DMWood}, and characteristic quantities 
such as dilatancy and internal friction angle are measured. Experimental data on elastic moduli~\cite{THHR90,Tat104,jia99,JM01,KJ02,GBDC03,SDH04,ARMJM05} are 
also extremely scarce below that range.
Most relevant laboratory sample histories to be understood in  order to relate the macroscopic response to internal 
variables and micromechanics involve an assembling stage, and then a 
compression stage, which is often isotropic or oedometric. It is therefore necessary to assess the influence of pressure changes on the initial
states. 

In addition, the material behavior under varying isotropic stress is interesting \emph{per se}. 
The behavior of sands is traditionally regarded~\cite{DMWood,GDV84,BiHi93} 
as \emph{elastoplastic}
under isotropic loading, with pressure cycles entailing irreversible density increases. Such effects are nevertheless considerably smaller than in cohesive materials such 
as clays~\cite{DMWood,BiHi93,MIT93}, or powders~\cite{Ca05}. It is worth investigating such behavior in model sphere packings by numerical means.
\section{Model material, micromechanical parameters}
\subsection{Contact model \label{sec:forces}}
We briefly recall here the model material and the contact laws, which are described in paper I with more details.
Equal-sized spherical beads of diameter $a$ (whose value, as we ignore gravity, will prove irrelevant),
interact in their contacts by point forces of elastic, frictional and viscous origins. 
The Hertz law relates the normal elastic force $N$ to the normal deflection $h$ (approach of sphere centers closer than $a$) as~:
\be
N=\frac{\tilde E\sqrt{a}}{3}h^{3/2},
\label{eqn:hertz}
\ee
with the notation ${\disty \tilde E = \frac{E}{1-\nu^2}}$, 
$E$ being the Young modulus of the beads, and $\nu$ the Poisson ratio. The Hertz law introduces a normal stiffness $K_N =  \frac{dN}{dh} $ that 
depends on $h$ or on $N$.

Tangential elasticity and friction are described with a simplified form of the
Cattaneo-Mindlin-Deresiewicz results~\cite{JO85}, in which the tangential stiffness $K_T$, relating the tangential elastic force increment
to the relative tangential elastic displacement $d{\bf u}_T$ in the contact,
is proportional to $K_N$:
\be
K_T= \frac{d{\bf T}}{d{\bf u}_T} = \alpha_T K_N\ \mbox{ with }\ \alpha_T =\frac{2-2\nu}{2-\nu} 
\label{eqn:tang}
\ee
The Coulomb condition with friction coefficient $\mu$ requires
${\bf T}$ to be projected back onto the circle of radius $\mu N$
in the tangential plane whenever the increment
given by Eqn.~\eqref{eqn:tang} would cause its magnitude to exceed this limit. 
In order to avoid unphysical increases of elastic energy, ${\bf T}$ is scaled down 
in proportion with $K_T$ when the elastic normal force $N$ decreases, as indicated in paper I and advocated in~\cite{EB96}.
Tangential contact forces also move with the particles in contact, so that the condition of objectivity is satisfied (see paper I and
ref.~\cite{KuCh06}).

A viscous term opposing normal relative displacements reads (positive normal forces are conventionally repulsive):
\be
N^v = \alpha(h) \dot h,
\label{eqn:fvisc}
\ee
with a damping coefficient $\alpha$ depending on elastic normal deflection $h$ (or on elastic repulsive force $N$), such that
its value is a fixed fraction $\zeta$ of the critical damping coefficient of the normal (linear) 
spring of stiffness $K_N(h)$ joining two beads of mass $m$:
\be
\alpha(h) = \zeta \sqrt{2mK_N(h)}.
\label{eqn:defzeta}
\ee
We do not introduce any tangential viscous force, and impose the Coulomb inequality to elastic force components
only. The main justification of such a term is computational convenience (to accelerate the approach of equilibrium states),
and we could check that its value did not affect the statistical results on the configurations of the packings.

The present numerical study was carried out with the elastic parameters
$E=70\,\mbox{GPa}$ and $\nu=0.3$ that are suitable for glass beads, and the friction coefficient
is set to $\mu=0.3$. 

\subsection{Stress control\label{sec:bcsc}}
The numerical results presented below were obtained on samples of $n=4000$ beads,
enclosed in a cubic or parallelipipedic cell with
periodic boundary conditions. The sizes of the cell 
are denoted as $L_\alpha$, parallel to coordinate axes $\alpha$ ($1\le \alpha\le 3$). 
$L_\alpha$'s vary simultaneously with the grain positions and orientations until mechanical equilibrium of
all particles with the prescribed values $\Sigma_{\alpha}$
of all three diagonal components $\sigma_{\alpha \alpha}$ of the Cauchy stress tensor, $1\le \alpha\le 3$, is obtained.
One then has~:
\be
\Sigma_{\alpha}= \frac{1}{\Omega}\left[\sum_i m_i v_i^\alpha v_i^\alpha +
\sum_{i<j} F_{ij}^{(\alpha)} r_{ij} ^{(\alpha)}
\right]
\label{eqn:stress}
\ee
Here $\Omega = L_1L_2L_3$ is the sample volume, $r_{ij}^{(\alpha)}$'s are the coordinates of vector ${\bf r}_{ij}$ joining the center of 
bead $i$ to the one of its contacting neighbor $j$ (with the nearest image convention of periodic cells) and $F_{ij}^{(\alpha)}$'s are those of the
corresponding contact force. This force is actually exerted by $i$ onto $j$, so that the convention used is that tensile stresses are negative.
Velocities ${\bf v}_i$ of grain centers comprise, in addition to a periodic field, an affine term corresponding to the global strain rate. Equations of
motion for dimensions $L_\alpha$ are written in addition to the ordinary equations for the dynamics of a collection of 
solid objects, and they drive the system towards an equilibrium state in which condition~\eqref{eqn:stress} is obeyed. 

In the present study we always impose isotropic stresses, \emph{i.e.} hydrostatic pressures $P$: $\Sigma_{\alpha} = P$ for $\alpha=1$, 2, 3.
\subsection{Dimensionless parameters\label{sec:control}}
In addition to include friction coefficient $\mu$ and viscous dissipation parameter $\zeta$,
the important dimensionless control parameters for  sphere packings under given pressure $P$ 
are the reduced stiffness $\kappa$ and the inertia parameter $I$.
$\kappa$ is chosen such that the typical contact deflection $h$ is proportional to $\kappa^{-1}$,
\be
\kappa= \left(\frac{\tilde E}{P}\right)^{2/3},\label{eqn:defkappa}
\ee
a correspondance which can be made accurate thanks to the relation
\be
P= \frac{z\Phi \ave{N}}{\pi a^2},
\label{eqn:relpn}
\ee 
between pressure $P=\tr \ww{\sigma}/3$ and the average normal force $\ave{N}$
in the contacts. 
\eqref{eqn:relpn} is exact provided $h\ll a$ in all contacts and intercenter distances are taken equal to the diameter $a$.
Here $z$ denotes the cordination number, equal to $z=2N_c/n$, with $N_c$ the total number
of force-carrying contacts in the packing. Rattlers, in proportion $x_0$, have no such contact. We refer to te force-carrying network - the
packing devoid of its rattlers -- as the \emph{backbone}, and to $z^*$, which simply relates to $z$ as r$z=(1-x_0)z^*$,  as the backbone coordination number.
Brackets denoting averages over all force-carrying contacts, one has
$$
\frac{\ave{h^{3/2}}}{a^{3/2}}=\frac{\pi}{z\Phi\kappa^{3/2}}.
$$
The limit of rigid grains is approached as $\kappa \to \infty$. 

$\kappa$ can be used to determine whether the material within the grains is likely to be imposed stresses beyond its elastic limit.
The maximum pressure, at the center of a Hertzian contact between spheres of diameter $a$, carrying a normal force $N$, is
\cite{JO85}
$$
p_{\text{max}}= \frac{2\times 3^{1/3}}{\pi} \frac{\tilde E^{2/3}}{a^{2/3}} N^{1/3}.
$$
Under pressure $P$, corresponding to $\kappa$ by~\eqref{eqn:defkappa}, when the average normal force in contacts is $\ave{N}$,
one can deduce from~\eqref{eqn:relpn}
\be
\frac{p_{\text{max}}}{\tilde E}= \frac{2\times 3^{1/3}}{\pi^{2/3}(z\Phi)^{1/3}} 
\left( \frac{N}{\ave{N}}\right)^{1/3} \kappa ^{-1/2}.\label{eqn:pmax}
\ee
Likewise, the maximum shear stress $\tau_{\text{max}}$, which is reached inside the grains near the contact region will be~\cite{JO85} (for $\nu=0.3$)
\be
\frac{\tau_{\text{max}}}{\tilde E}= 0.31 \frac{p_{\text{max}}}{\tilde E}.\label{eqn:taumax}
\ee
Eqns.~\ref{eqn:pmax} and~\ref{eqn:taumax} show that very high stress levels, up to a non-negligible fraction of elastic modulus
$E$ are reached if $\kappa$ is not large enough. With our choice of material parameters for glass beads, 
we get $\kappa^{-1/2}\simeq 0.051$ for P=10~MPa and $\kappa^{-1/2}\simeq 0.11$ for P=100~MPa, while the numerical prefactor is only
slightly lower than 1 ($\sim 0.8$) if $z\Phi = 4$ (a typical value) in~\eqref{eqn:pmax}. Such high stresses are very likely to entail particle
breakage or plastic strains (according to the materials the grains are made of).

In our simulations we set
our lowest pressure level for the simulation of glass beads to 1~kPa or 10~kPa, corresponding
to $\kappa \simeq 181000$ and $\kappa \simeq 39000$ with the
elastic properties of glass. This enables us to explore  the entire experimental pressure range, and to approach the large $\kappa$ limit too.
Up to the maximum pressure value 100~MPa, we assume elastic contact behavior, but
one should be careful on comparing the numerical results in the higher pressure states ($P\ge 10$~MPa) to 
experimental ones.

Dynamical effects are assessed on comparing the strain rate $\dot \epsilon$ to intrinsic inertial times, such as the time needed for
a particle of mass $m$, initially at rest, 
accelerated by a typical force $Pa^2$, to move on a distance $a$. This leads to the definition of 
a \emph{dimensionless inertia parameter} :
\be
I=\dot \epsilon\sqrt{m/aP}.\label{eqn:defI}
\ee
The quasistatic limit can be defined as $I\to 0$. $I$ is a convenient parameter to describe internal states 
and write down constitutive laws for
granular materials in dense shear flow~\cite{GdR04,Dacruz05,PouliquenPG,JoFoPo05,JFP06}.
\subsection{Initial states}
\begin{table*}
\caption{Isotropic states ($\kappa\simeq 39000 $ for A and C, $\kappa\simeq 181000$ for B and D) for different assembling procedures.}
\centering
\begin{tabular}{|l|cccccc|}  \cline{1-7}
Procedure & $\Phi$ & $z^*$ & $x_0$ (\%) &Z(2) & $M_1$ &$M_2$\\
\hline
\hline
A &  $0.6370\pm 0.0002$ &$6.074\pm 0.0015$ & $1.3\pm 0.2$&$1.53$&0&0\\
\hline
B ($\mu_0=0.02$) &$0.6271\pm 0.0002 $ &$5.80\pm  0.007$  & $1.95\pm 0.02 $&$1.52$&$0.016$&$0.018$\\
\hline
C (vibration) &$0.635\pm 0.002 $ &$4.56\pm  0.03$  &$13.3 \pm 0.5$&$1.65$&$0.135$&$0.181$\\
\hline
D &$0.5923\pm 0.0006$ &$4.546\pm 0.009$ & $11.1\pm 0.4$&$1.58$&$0.160$&$0.217$\\
\hline
\end{tabular}
\label{tab:prep}
\normalsize
\end{table*}
The present paper is devoted to the study of the influence of quasistatic pressure changes to granular packings assembled by different means,
as described in paper I~\cite{iviso1}. Four different states were prepared under low pressure, and some of their basic characteristics are recalled 
in table~\ref{tab:prep}. 
Such state variables are monitored in the following as a function of pressure in isotropic compression or pressure cycles. In addition to
solid fraction $\Phi$, proportion of rattlers $x_0$, backbone (or force-carying structure) coordination number $z^*$, Table~\ref{tab:prep} 
provides some global information on force distributions. $Z(2)$ is characteristic of the width of the distribution of normal forces:
\be
Z(2) = \frac{\ave{N^2}}{\ave{N}^2}.
\label{eqn:defza}
\ee
$M_1$ and $M_2$ are the average levels of friction mobilization (\emph{i.e.,} ${\disty \frac{\norm{T}}{N}}$) for contacts
carrying normal forces, respectively, larger and smaller than the average $\ave{N}$.

In paper I we also recorded other geometric data, in particular pair correlation functions and distributions of near neighbor gaps $h$. The latter can be expressed
as gap-dependent coordination numbers, defining $z(h)$ as the average number of neighboring beads around a central one, separated by an interstice
smaller than $h$. $z(0)$ thus coincides with the contact coordination number. Due to the rattlers, the proportion of which --see table~\ref{tab:prep}--  can
exceed 10\% of the total number of grains, such geometric data are however somewhat ambiguously defined: the positions of the rattlers are not fixed by the
rigid backbone. Thus one may define $z^I(h)$, on using the arbitrary positions obtained at the end of the simulation, when the packing first equilibrates within the
prescribed numerical tolerance. One then has $z^I(0)\simeq z$ (recall $z$ counts only force-carrying contacts)
if the equilibrium state is accurately computed, because there are very few contacts bearing a normal force below tolerance. In an attempt to define more intrinsic
geometric data, we defined $z^{II}(h)$ in paper I~\cite{iviso1} as the gap-dependent coordination number in the configuration obtained once all rattlers are pushed
against the backbone, in random directions. In their new position, the rattlers now have three contacts with the backbone (except in the 
rare case when inter-rattler contacts are obtained). It was argued in paper I that the resulting structure was likely to resemble, to some extent, 
granular assemblies under gravity, when the weight of the grains is very small in comparison to the local stress. $z^{II}(0)$ can be regarded as a 
\emph{geometric} definition of a contact coordination number (it is, in general, slightly larger than $z^*=z/(1-x_0)$).
\section{Numerical results\label{sec:comp}}
We first specify the numerical compression procedure in paragraph~\ref{sec:compnum}, then describe the effects of an isotropic compression and a pressure cycle in terms of
global variables (Section~\ref{sec:compglob}) as well as local geometry (Section~\ref{sec:compmicro}). 
We then test the simplest prediction scheme for the evolution of
coordination number, that of homogeneous strain at the microscopic level, in Section~\ref{sec:zpred}.
\subsection{Numerical procedure\label{sec:compnum}}
The results presented below pertain to equilibrium configurations at variable isotropic pressure $P$, 
obtained by a stepwise compression (respectively: decompression) process
in which $P$, within the controlled stress scheme described in Section~\ref{sec:bcsc}, is increased (respectively: decreased) by a factor $\sqrt{10}$. 
In each pressure step a condition
of slow enough strain rate was enforced, so that the inertia parameter, as defined by~\eqref{eqn:defI} 
with the currently imposed pressure level, was kept below a maximum value: $I\le 10^{-3}$
for compression, $I\le 10^{-4}$ for decompression. Such values were chosen to ensure independence of the results on dynamical parameters $I$ and $\zeta$. 
It was observed that
a decompression process requested more care, due to its greater instability. 
Whereas a  compression of the sample beyond its equilibrium density will be strongly opposed, at growing $P$,
by elastic forces in the network, too large an expansion, as $P$ decreases, might cause the contact network to break apart, 
resulting in a dynamical process similar to assembling
a granular gas, when the externally applied pressure finally drives the system back to a denser equilibrium configuration. 
Such events might entail a significant remoulding of the contact network and large departures from equilibrium conditions. This 
should of course be avoided in a procedure designed to model a quasistatic evolution, as close as possible to the limit of small strain rates.

Configurations are deemed equilibrated when, defining $\epsilon_F = 10^{-4} Pa^2$ as a small tolerance
on forces and $\epsilon_E = 10^{-7} Pa^3$ as a small tolerance on energies, the four following conditions are simultaneously satisfied~:
\bi
\im
each coordinate of the total force on each grain is smaller than $\epsilon_F$;
\im
each coordinate of the total moment on each grain is smaller than $\epsilon_F a$; 
\im
all stresses have their prescribed values with a relative error smaller than $\epsilon _F$:
$$
(\alpha = 1,\ 2,\ 3)\ \ \frac{\vert\sigma_{\alpha\alpha} - P\vert}{P} < \epsilon _F
$$
\im 
the kinetic energy per grain is smaller than $\epsilon_E$.
\ei

To distinguish between the backbone and the rattlers, the same method is
applied as presented in paper I~\cite{iviso1}.

Such procedures were applied to samples A to D below, with $P$ ranging from its smallest value 1~kPa 
(for B and  D, corresponding too $\kappa\simeq 181000$),
or 10~kPa (for A and C, corresponding to $\kappa\simeq 39000$), up to 100~MPa ($\kappa \simeq 84$), 
and then back to its initial low value. Letters A, B, C, D will hereafter
denote pressure-dependent configuration \emph{series}.
Although initial states A and B were assembled with
coefficients of friction lower than the chosen value $\mu=0.3$, we study quasistatic compressions with $\mu=0.3$ for all sample series. 
We regard the smaller friction levels applied to configurations A and B in the assembling stage
as models for lubricated grains, and assume that the lubricant ceases to operate once solid particles finally touch one another, 
as in equilibrated packings and during quasistatic compression tests. As a reference for 
comparisons with other states, and because it was studied in the literature~\cite{MJS00,Makse04}, 
we also prepared another configuration series we denote as A0, obtained from the
initial A state on compressing a frictionless system (thus series A and A0 share the same initial low-pressure state, but differ as soon as $P$ is altered). 

All results are averaged over 5 samples of n=4000 beads, and error bars correspond to one standard deviation.
\subsection{Evolution of global state variables\label{sec:compglob}} 
\begin{figure}[htb]
\includegraphics*[angle=270,width=8.5cm]{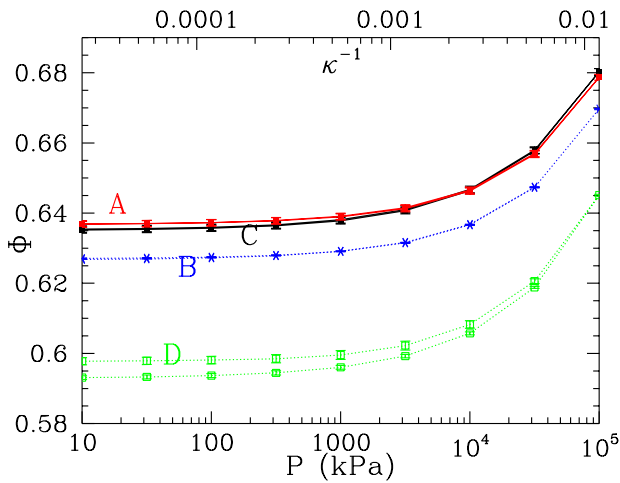}
\caption{\label{fig:phip}
(Color online) Evolution of packing fraction as a function of pressure $P$ in glass bead packings (bottom axis), 
or dimensionless stiffness parameter $\kappa^{-1}$ (top axis), 
in (from top to bottom) states A (red crosses, continuous line), C (black square dots, continuous line) 
B (blue asterisks, dotted line) and D (green open squares, dotted line). 
}
\end{figure}

\begin{figure}[htb]
\subfigure[$z^*$ versus $P$ or $\kappa^{-1}$]{
\includegraphics*[angle=270,width=8.5cm]{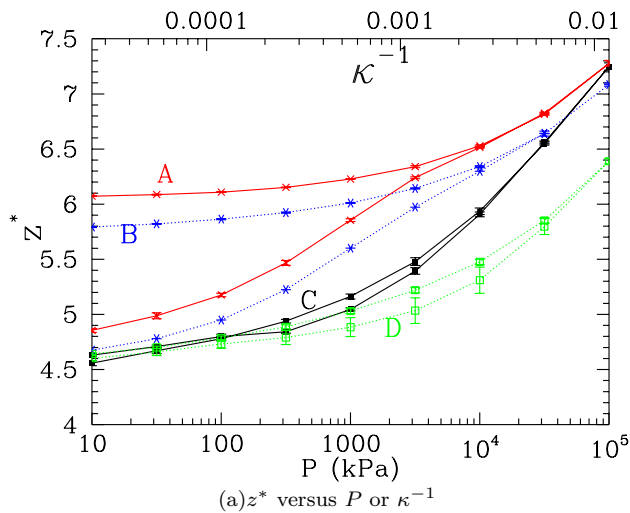}
\label{fig:zp}}
\subfigure[$x_0$ versus $P$ or $\kappa^{-1}$]{
\includegraphics*[angle=270,width=8.5cm]{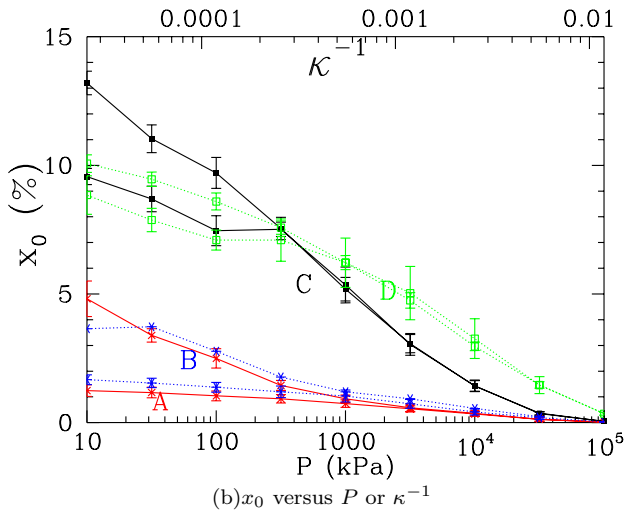}
\label{fig:x0p}}
\caption{\label{fig:zpx0p}
(Color online) Backbone coordination number $z^*$ (a) and proportion of rattlers $x_0$ (b) as functions of $P$ or $\kappa^{-1}$, 
same symbols as on Fig.~\ref{fig:phip}.}
\end{figure}

Figs.~\ref{fig:phip} and \ref{fig:zpx0p} display the evolution of solid fraction $\Phi$, backbone coordination number $z^*$, and
rattler fraction $x_0$ in sample series A, B C and D in the pressure cycle. 
Fig.~\ref{fig:phip} shows that the solid fraction change with pressure is almost perfectly reversible: the data points corresponding to the compression and decompression
parts of the pressure cycle are almost indistinguishable. More precisely, once the pressure had returned to its 
lowest value in samples A to C, the packing fraction was observed to have changed by very small amounts, below $2\cdot 10^{-4}$. 
The loosest state, D, undergoes a slight compaction. Yet, this effect apparently decreased as
the maximum prescribed value for parameter $I$ was changed from $10^{-4}$ to $10^{-5}$ upon unloading (the reported results corresponding to this latter value). 
Our model material thus differs from sands, which are reported to respond to such cycles with notable irreversible 
density increases~\cite{BiHi93,DMWood}. It should be noted, though, that we are using a contact model without plasticity or particle damage, 
which, as argued on evaluating, in Sec.~\ref{sec:control},
the maximum pressure and shear stress in the grains near contact points with 
Eqns.~\eqref{eqn:pmax} and \eqref{eqn:taumax}, is quite unrealistic for the highest pressure
levels simulated. Stress concentrations
in contacts between angular particles like sand grains, with corners or asperities~\cite{JO85,GO90}, are more severe than between
smooth objects and should enhance the effects of anelastic material behavior within the grains.
The smallness of irreversible compaction in our simulations suggests that such macroscopic behavior, in sands, 
originates in contact mechanics rather than in collective 
effects.

The reversibility of the response to the pressure cycle is however only apparent, 
as the coordination number does not return to its initial value. 

As expected, $z^*$ increases under
a growing confining pressure (Fig.~\ref{fig:zp}): as the particles pack more closely in a smaller volume, near neighbors come into contact. 
$z^*$ reaches about 7.3 at the highest pressure in the densest samples, A and C. 
Correlatively, an increasing number of rattlers get trapped as their free volume shrinks, and are recruited by the force-carrying network. 
The initially large fraction of 
rattlers in states C and D ($x_0 > 10\%$) steadily decreases as $P$ grows( Fig.~\ref{fig:x0p}) and has virtually disappeared at $P=100$~MPa. 

\begin{figure}[!htb]
\includegraphics*[angle=270,width=8.5cm]{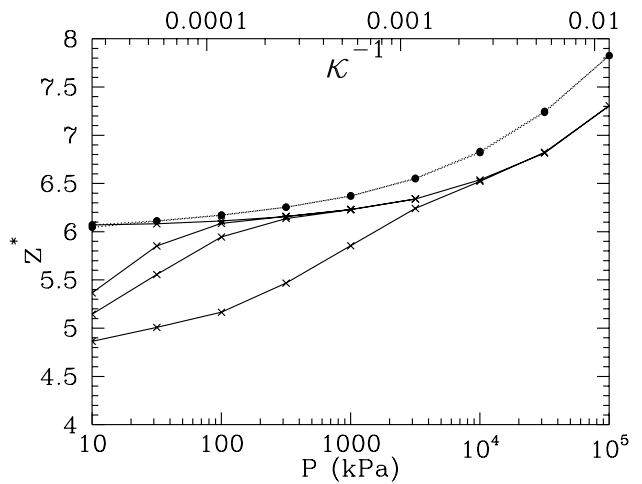}
\caption{\label{fig:zpA}
 $z^*$ versus $P$ or $\kappa^{-1}$ in pressure cycle in series A (crosses) and A0 (dots), showing reversibility for A0.
Shorter cycles (up to 0.316~MPa and 1~MPa) than the one of Fig.~\ref{fig:zpx0p} are also shown for A.
}
\end{figure}

\begin{figure}[htb]
\includegraphics*[width=8.5cm]{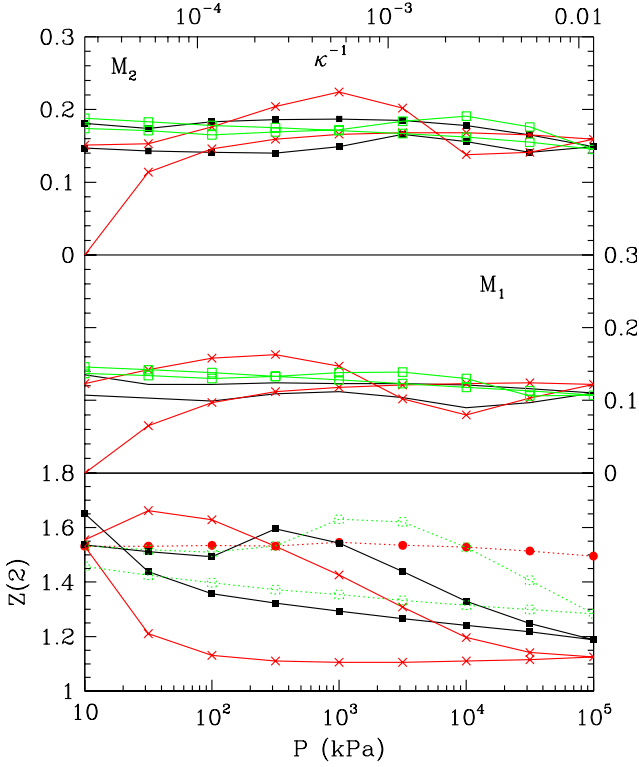}
\caption{\label{fig:pdisp}
(Color online) From bottom to top: $Z(2)$, $M1$ and $M2$ versus $P$ or $\kappa^{-1}$ in compression cycle. 
Symbols as on Fig.~\ref{fig:phip} for states A, C, and D. 
Series A0 represented
with (red) dots joined by dotted line for $Z(2)$. Hysteresis loops for $Z(2)$ first decrease, 
then increase back on unloading and go through a maximum (except for A0, in which cas it
is nearly constant). $M_1$ and $M_2$ behave in a similar way, with the special circumstance that their initial values are equal to zero in A states 
(assembled without friction).
}
\end{figure}

\begin{figure}[htb]
\includegraphics*[width=8.5cm]{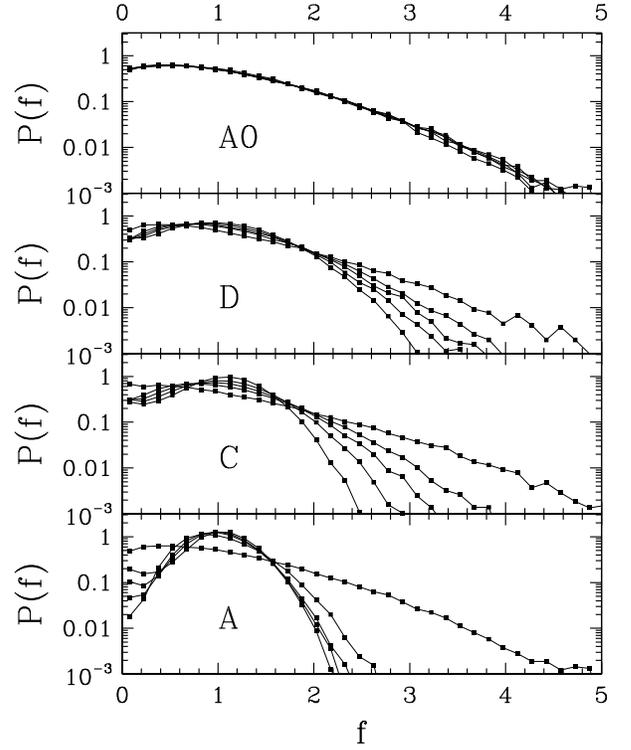}
\caption{\label{fig:histp}
From bottom to top, evolution of normalized force distributions $P(f)$, with $f=N/\ave{N}$,
 with \emph{growing} pressure in samples A, C, D, A0. $P$ value in kPa are $10$ (except for D: $P=1$), $100$, $10^3$, $10^4$ and $10^5$. 
All four distributions tend to narrow as $P$ grows, but at very different rates. 
}
\end{figure}

The evolution of coordination numbers on \emph{unloading} is more surprising. 
While low coordination states C and D exhibit a very limited hysteresis effect and eventually
retrieve their initial, low $z^*$ values (about $4.6$), with a slightly lower rattler fraction, samples of types A and B, 
in which $z^*$ was initially high, lose contacts as
a result of the pressure cycle and end up with $z^*$ values below 5 (about 4.8 for A, and 4.5 for B), 
closer to C and D ones than to where they started, with a substantial rise in the
population of rattlers. (Let us recall that samples A and B are regarded in the study of quasistatic compression 
as made of frictional beads with $\mu=0.3$, like the others). The behavior of 
(frictionless) samples A0 is of course different, for they cannot be stable at low pressure below $z^*=6$~\cite{JNR2000}.
Fig.~\ref{fig:zpA} compares the evolutions of $z^*$ in A and A0 series, and shows that
$z^*$ is very nearly reversible in the A0 series. The unloading curves in A states starting at lower pressures, 
3.16~Mpa and 1~MPa instead of 100~MPa, also shown on Fig.~\ref{fig:zpA}, witness a lower,
but significant decrease of $z^*$ from its initial value $z^*\simeq 6$ at the end of the cycle. 
The shape of the force distribution and the mobilization of friction 
also change with $P$, as shown by the evolution of parameters $Z(2)$, $M_1$, $M_2$ on Fig.~\ref{fig:pdisp}.
As a general rule, the width of the force distribution correlates with the level of force indeterminacy, relatively to the number of degrees of freedom. 
Contact elasticity tends to
share forces rather evenly, because contact force values should minimize the intergranular elastic energy, subject to the constraint that they balance the applied 
pressure (this elastic energy as a function of forces is written further below in connection with a discussion of irreversibility in pressure cycles, 
and the minimization
property is exploited in paper III~\cite{iviso3} to estimate bulk moduli).
More precisely, the \emph{increments} of forces due to pressure increases will tend 
to reduce the width of the distribution, the faster the less constrained the minimization,
\emph{i.e.} the larger the degree of force indeterminacy. Thus in configurations A, the large coordination number enables
a quick narrowing of the distribution under growing pressure. In states C, the same tendency is present, but the evolution is much slower, 
as there are less possibilities to
distribute forces in a more tenuous network while maintaining equilibrium. However, C samples gain contacts faster than D ones (Fig.~\ref{fig:zp}), for which the
narrowing effect is even slower. Finally, the extreme case is the situation of isostaticity, as in the A0 series, 
in which the distribution of forces is geometrically determined in the rigid limit of $\kappa\to +\infty$.
As, furthermore, the increase of $z$ with $P$ is not very fast in that case, since $z$ is already large from the beginning, 
the shape of the distribution remains nearly 
constant. A few normal force probability distribution functions at different pressure levels are shown on Fig.~\ref{fig:histp}. 

\begin{figure}
\includegraphics*[angle=270,width=8.5cm]{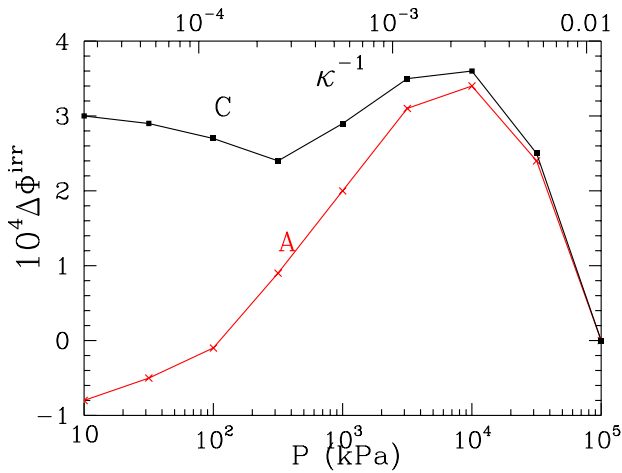}
\caption{(Color online) Increment of packing fraction $\Delta \Phi^{\text{irr}}$ gained between the two states of equal pressure, reached at 
growing and at decreasing $P$, in states A (red, crosses) and C (black, square dots). Note the scale of density changes ($\Delta \Phi$ of order $10^{-4}$). 
\label{fig:phidiff}
}
\end{figure}

The evolution of force values and friction mobilization on unloading is more complicated: all three parameters shown on 
Fig.~\ref{fig:pdisp} first increase, then go through a maximum
and end up, at the initial pressure value, with a value comparable with the initial one (except for friction mobilization parameters $M_1$ and $M_2$ in A systems, 
because they  started at zero). 
In a granular sample controlled in displacements or strains, rather than stresses, large self balanced forces can in some situations 
remain when the external load that created them
is removed, the simplest example being that of one particle wedged in a corner~\cite{McGaHe05,HaEr99}. 
Our observations indicate that such a phenomenon does not take place in a situation
of controlled stress state: all forces are of the order of the average force, which is related to the current pressure by~\eqref{eqn:relpn}, 
even though contacts have carried forces that were larger by orders of
magnitude in the past. This suggests that the set of admissible contact forces, 
restricted  to the intersection of an $h$-dimensional affine space (due to
equilibrium relations) with a cone (due to Coulomb inequalities) is bounded. 
Yet during unloading many more sliding contacts are observed than at growing pressure, due to the effects
of decreasing normal force components, and the level of friction mobilization is higher (Fig.~\ref{fig:pdisp}). 
Meanwhile, the distribution of normal forces gets wider. The global influence of
the past loading, with contacts previously carrying larger forces, enhances force heterogeneities.
A related quantity is the \emph{elastic energy} stored in the contacts. The total elastic energy per grain $w$ reads (from Eqns.~\eqref{eqn:hertz} and
\eqref{eqn:tang})
$$
w = \frac{1}{n} \sum_{i=1}^n \sum_{j\ne i} \left[\frac{3^{2/3}}{5\tilde E^{2/3}a^{1/3}}N_{ij}^{5/3} + \frac{\normm{{\bf T}_{ij}}^2}{4K_T(N_{ij})}\right].
$$
 Once adimensionalized by $\tilde E a^3$, we denote it as $\tilde w$. 
On exploiting Eqn.~\eqref{eqn:relpn} it is conveniently expressed as:
\be
\tilde w = \frac{3^{2/3}\pi^{5/3}}{5}\frac{\tilde Z(5/3)}{z^{2/3}\Phi^{5/3}\kappa^{5/2}}.
\label{eqn:tilw}
\ee
In \eqref{eqn:tilw}, $\tilde Z(5/3)$, related to force moments, is close to $Z(5/3)$, which can be defined on replacing exponent
2 by $5/3$  in Eqn.~\eqref{eqn:defza}, with the following slight modification.
With $\alpha_T$ defined in~\eqref{eqn:tang} as the constant ratio of tangential to normal stiffnesses, 
 and with the notation $r_{TN}$ for the
ratio ${\disty \frac{\norm{T}}{N}}$ in a contact, let us define
\be 
\tilde Z(5/3) = \frac{\langle N^{5/3}(1+\frac{5r_{TN}^2}{6\alpha_T})
\rangle }{\langle N\rangle ^{5/3}}.
\label{eqn:tilz}
\ee
$\tilde Z(5/3)$ thus depends on the force distribution and also on friction mobilization, although for $\mu=0.3$  its relative difference with $Z(5/3)$ 
is small (of the order of $\frac{5M_1^2}{6\alpha_T}$, with $M_1$ as plotted on Fig.~\ref{fig:pdisp}). 
The energy per particle, $\tilde w$, scales as $\kappa^{-5/2}$, which is
expected since this is proportional to $h^{5/2}$ for $h\propto \kappa ^{-1}$ the typical normal contact deflection. 
$\tilde w$ is larger for low coordination numbers (weaker
networks), and larger force disorder (higher $\tilde Z(5/3)$). (It should be recalled that we use pressure, rather than strain, as the control parameter, hence
a larger elastic energy for softer materials). Thus in A configurations, $\tilde w$ is larger, for given $\kappa$, on \emph{decompressing}, another
manifestation of the irreversibility of the cycle. 
If we assume that the curve $P(\Phi)$ is quasistatically followed up to the maximum pressure, and then exactly retraced back on
decompressing, this leads to a paradox, as some elastic energy appears to be gained at no expense. 
Thus one has to account for very small irreversible density changes, for energetic
consistency. Such changes in $\Phi$, between the growing and decreasing pressure parts of the cycle are shown on Fig.~\ref{fig:phidiff}.
In the case of A configurations, one even observes a slight \emph{decompaction} on decreasing $P$ back to its lowest, initial value. Although
surprising, this phenomenon should be expected in the rigid limit $P\to 0$ or $\kappa \to \infty$, because as explained in paper I, the initial
A configuration, which was assembled without friction, is a local maximum of $\Phi$ subject to impenetrability constraints. 
Another conclusion of paper I~\cite{iviso1}
is that the only way to increase density in such a sample is to produce, by enduring agitation or repeated shakes, 
notable traces of crystalline order. This should not 
happen in a slow, quasistatic compression experiment with only one pressure cycle.
To check for energetic consistency, one may
note on Fig.~\ref{fig:phidiff} however that the change of $\Phi$ is positive at high pressure. The total energy fed into the system in the cycle is
\be
\Delta\tilde w ^{\text{ext}}= \frac{\pi}{6}\int \frac{\Delta \Phi^{\text{irr}}(P) dP}{\tilde E \Phi^2},
\label{eqn:dwext}
\ee
the integral running over the whole pressure interval of the compression cycle.
Consequently (see Fig.~\ref{fig:phidiff}) the contribution of the irreversible 	\emph{increase} 
of $\Phi$ is largely dominant, because it is integrated over a
much wider pressure interval.
The small changes in density between the compression and the decompression
curves at the same pressure values are large enough to explain the change
in elastic energy, and that of potential energy as well when the cycle ends up \emph{decreasing} the density (which happens for A samples). 
\subsection{Pair correlations and near neighbor distances\label{sec:compmicro}}
The smallness or absence of irreversible compaction in the pressure cycle implies that the samples do not avoid contact deflections
by finding denser packing arrangements. Thus interparticle correlation patterns should witness favored near neighbor distances
which typically scale like $\Phi^{-1/3}$.
\begin{figure}[htb]
\subfigure[$g(r)$ versus $r/a$ in C configurations at different $P$.]{
\includegraphics*[angle=270,width=8.5cm]{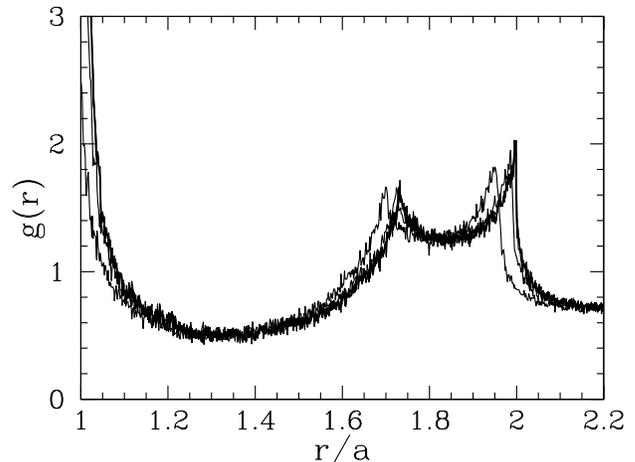}
\label{fig:grphiQ0}}
\subfigure[$g(r)$ versus $r^*/a$ in C configurations at different $P$.]{
\includegraphics*[angle=270,width=8.5cm]{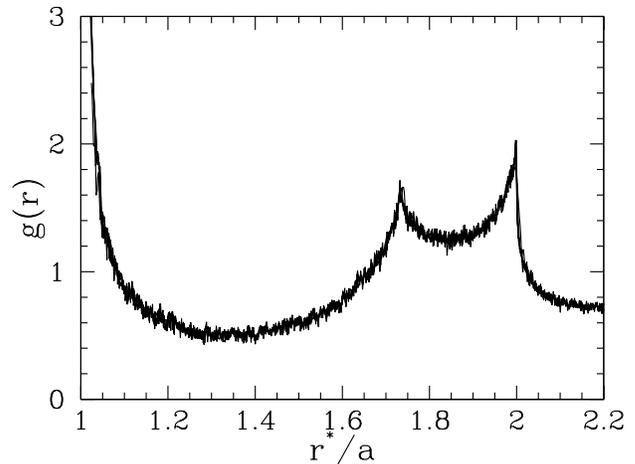}
\label{fig:grphiQr}}
\caption{\label{fig:grphiQ}
Pair correlation functions at P=10, $100$, $1000$, $10^4$, $10^5$~kPa in configurations C at growing pressure, without (top), and with (bottom)
rescaling distance $r$ as $r^* = r (\Phi/\Phi_0)^{1/3}$.}
\end{figure}
This is shown for C configurations on Fig.~\ref{fig:grphiQ}: on rescaling the distance axis, 
using coordinate $r^* = r (\Phi/\Phi_0)^{1/3}$ with $\Phi_0$ the initial
low pressure solid fraction, 
the different $g(r)$ curves are superimposed. In agreement with the observations made in paper I~\cite{iviso1}, where the relationships between
pair correlation functions and contact networks were discussed, 
a closer look on such correlations will reveal
differences in the details of the peaks associated with changes in the coordination number with $\Phi$.
Figs.~\ref{fig:zhxpPQE} and~\ref{fig:rzhxpPQE} respectively show functions $z^I(h)$ and $z^{II}(h)$ at growing $P$ values, 
using the corresponding change of scale for interstice $h$, $h^*= (\Phi/\Phi_0)^{1/3}(a+h)-a$.
\begin{figure}[htb!]
\includegraphics*[angle=270,width=8.5cm]{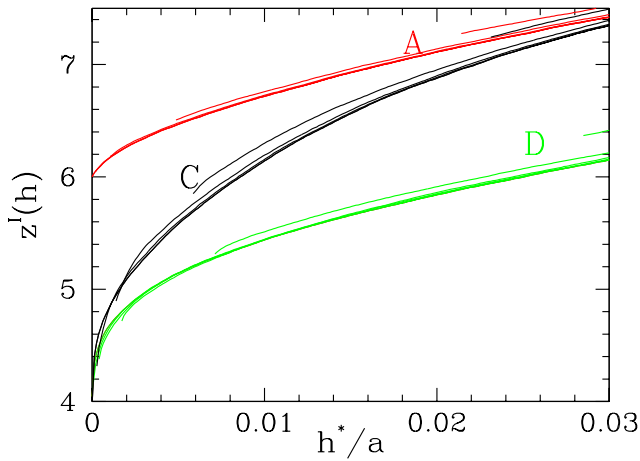}
\caption{\label{fig:zhxpPQE}
(Color online) Gap-dependent neigbor coordination number $z^I(h)$ 
versus rescaled interstice $h^*= (\Phi/\Phi_0)^{1/3}(a+h)-a$ at different $P$ (same as on Fig.~\ref{fig:grphiQ})
in states A (red), C (black) and D ( green).}
\end{figure}
\begin{figure}[htb!]
\includegraphics*[angle=270,width=8.5cm]{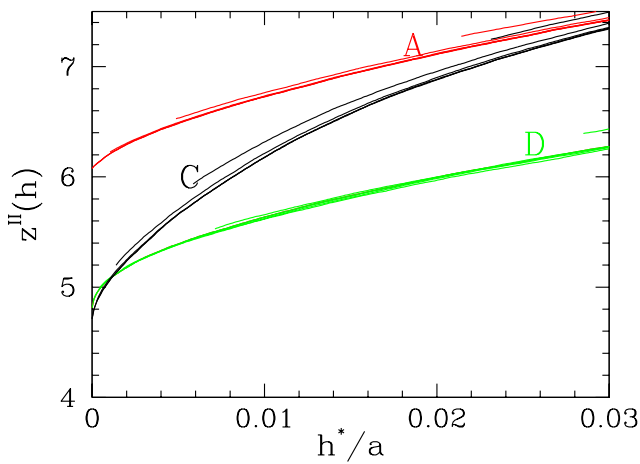}
\caption{\label{fig:rzhxpPQE}
(Color online) Same as Fig.~\ref{fig:zhxpPQE} for definition $z^{II}(h)$ of the gap-dependent neigbor coordination number.}
\end{figure}
Those data suggest that the homogeneous shrinking of distances implied by the 
rescaling of abscissae on the graphs of Figs.~\ref{fig:grphiQr}, \ref{fig:zhxpPQE} and 
\ref{fig:rzhxpPQE} is an approximation with some discrepancies at small intergranular distances. 
Curves corresponding to pressures other than the lowest one on
Figs.~\ref{fig:zhxpPQE} and \ref{fig:rzhxpPQE} start at distance $[(\Phi/\Phi_0)^{1/3}-1]a>0$ and the corresponding values of 
$z(h)$ on the curve for the lowest pressure value are the predictions for the coordination
number on assuming homogeneous shrinking strains. Differences therefore show that such predictions, 
albeit reasonable, are not exact. In particular, the gradual
capture of rattlers by the force-carrying network as $P$ grows (see Fig.~\ref{fig:x0p}) 
cannot be adequately described by the homogeneous shrinking assumption: the rattlers
will not start carrying forces when one interstice with a backbone grain is closed. 
The use of definition $z^{II}(h)$ should in principle improve this kind of prediction:
once positioned  against the backbone (with 3 contacts), the rattlers are 
much more likely to create new contacts bearing nonzero forces when they touch new neighbors. 
Yet, the improvement of curve superpositions on Fig.~\ref{fig:rzhxpPQE} compared to
Fig.~\ref{fig:zhxpPQE} is marginal. This suggests that the inaccuracy of 
the prediction of coordination numbers  is not only due to the capture of rattlers by the growing backbone, but also stems from
the failure of the assumption of homogeneous shrinking.
\subsection{Can one predict the changes in coordination number ?\label{sec:zpred}} 
\begin{figure}[htb!]
\includegraphics*[angle=270,width=8.5cm]{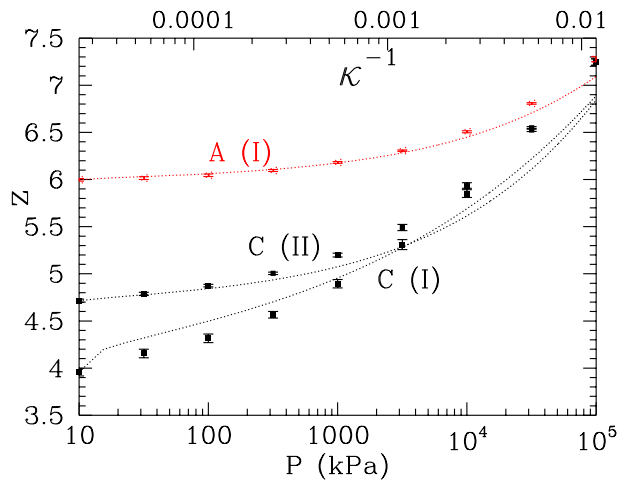}
\caption{\label{fig:zppred}
(Color online) Predictions for $z=z^I(0)$ in samples A and C, and for $z^{II}(0)$ in samples C, based on the homogeneous shrinking assumption.}
\end{figure}

\begin{figure}[htb!]
\includegraphics*[angle=270,width=8.5cm]{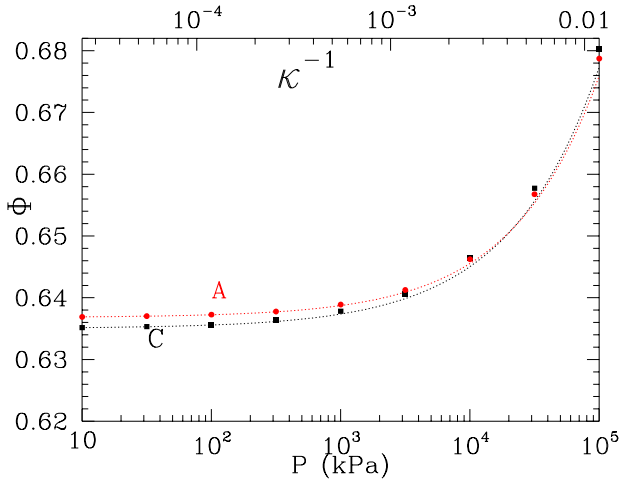}
\caption{\label{fig:phiplinpred}
(Color online) $\Phi$ versus $P$ or $\kappa^{-1}$ in samples A (red) and C (black). Dots: measurements. Dotted lines: predictions, 
based on the homogeneous shrinking assumption from the initial state of lowest pressure. }
\end{figure}

The results of the prediction of the coordination number, assuming all distances uniformy shrink, 
are shown on Fig.~\ref{fig:zppred} for systems A and C under growing pressure. The
agreement is very good in state A (except at high pressure, where $z$ is slightly underestimated), and fair in state C. For C configurations, 
the prediction was done separately for both $z$ and $z^{II}(0)$, showing a somewhat better accuracy at low pressure in the second case.
Unfortunately, the mechanically important coordination number is $z^I(0)=z$ rather than $z^{II}(0)$.
To evaluate $P$ as a function of $\Phi$, one needs to account for two phenomena: the increase of the elastic normal deflection 
in the contacts that already existed at the lowest pressure, 
and the creation of new contacts due to the closing of open interstices. 
Both effects are evaluated with the assumption of homogeneous rescaling of all distances according to the
density change, respectively exploiting the previous measurements of the 
distribution of sphere overlaps (related to that of normal forces), and of the function $z(h)$ 
(with no significant difference in accuracy on using $z^I$ or $z^{II}$). The predicted 
values of $z$, although not very accurate for small changes of $z^I$ at low pressures, globally capture the marked growing trend above 1~MPa. 
The predictions of density increases are compared with the simulation results on Fig.~\ref{fig:phiplinpred}, showing good agreement (with a slight
underestimation at high pressure). The prediction of $P$ is understandably more accurate than that of the coordination number, 
because it is not very sensitive, at first, to
errors in the estimation of the density of newly created contacts, which initially carry very small forces.

\begin{figure}[htb!]
\includegraphics*[angle=270,width=8.5cm]{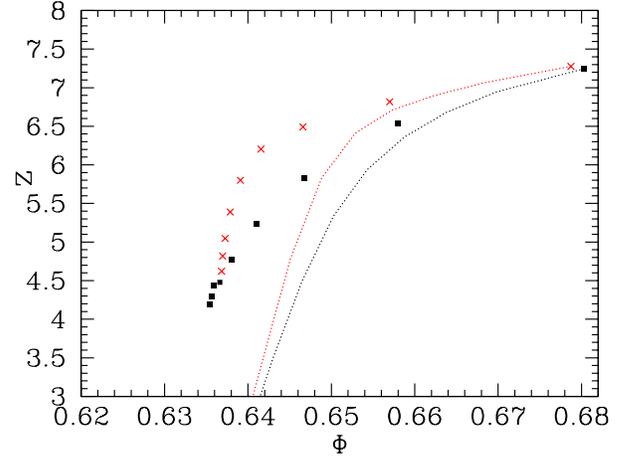}
\caption{\label{fig:rzphipred}
(Color online) Coordination number $z$ versus $\Phi$ at decreasing $P$ in samples A (red) and C (black). Dots: measurements. 
Dotted lines: predictions, based on the homogeneous expansion assumption from the
initial state of highest pressure.}
\end{figure}

\begin{figure}[htb!]
\includegraphics*[angle=270,width=8.5cm]{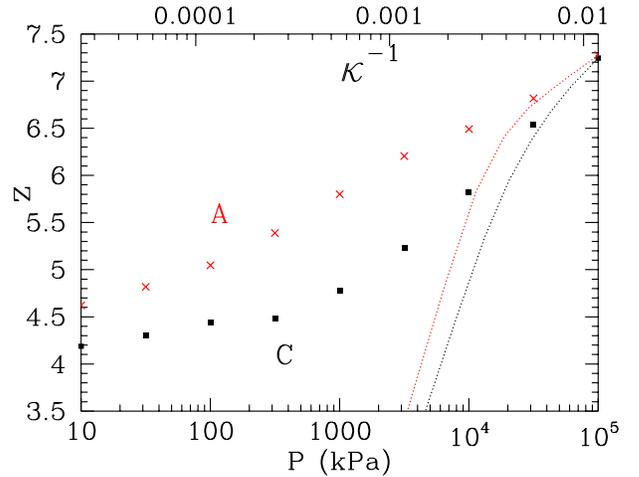}
\caption{\label{fig:rzppred}
(Color online) Coordination number $z$ versus decreasing $P$ (or $\kappa^{-1}$) in samples A (red) and C (black). Dots: measurements. 
Dotted lines: predictions, based on the homogeneous expansion assumption from the
initial state of highest pressure.}
\end{figure}

One may also attempt to predict the decrease of coordination number in the 
decompression part of the pressure cycle. Such a prediction is based on the distribution of 
particle overlaps (or contact deflections), rather than near neighbor distances. 
The relevant information is therefore the normal force histogram for the highest pressure level,
as shown, \emph{e.g.} on Fig.~\ref{fig:histp}. However, this is a rather crude approximation, 
which leads to large errors for the coordination number variation with
density, as shown on Fig.~\ref{fig:rzphipred}, and very poor predictions 
indeed for the coordination number relationship to the decreasing pressure, as apparent on
Fig.~\ref{fig:rzppred}. 
Such an assumption of homogeneous expansion proves in particular unable to provide a correct 
estimate of the properties at low density or pressure, as it ignores the requirement
of mechanical rigidity. 
We are not aware of a simple prediction scheme that would be able to provide a reasonably accurate 
description the reduction of coordination number in the A state on reducing the confining pressure.
\section{Discussion \label{sec:compdisc}}
The effect of a compression on the four series of isotropic packings we have been studying can 
be broadly summarized as the closing of additional contacts and
the gradual reduction of the characteristic disorder of granular systems, as witnessed by the narrowing
of the force distribution (Figs.~\ref{fig:pdisp} and \ref{fig:histp}). 
Geometric changes conform to the homogeneous shrinking assumption on large scale, and the resulting predictions for the near-neighbor distances
and the coordination numbers are reasonable, if not very accurate,
approximations (Figs.~\ref{fig:zppred} and \ref{fig:phiplinpred}), 
even though they cannot correctly account for the recruitment of rattlers (Fig.~\ref{fig:x0p}) by the growing backbone. It proves difficult 
to accurately estimate small $z^*$ increases, to which, as will be studied in~\cite{iviso3} (paper III), shear moduli
of poorly coordinated packings are especially sensitive. The changes in the forces and 
the mobilization of friction are not appropriately described by such a simple model. On assessing the performance of the homogeneous shrinking approximation,
one thus retrieves the classification of length scales introduced in paper I~\cite[Section IV.E.2]{iviso1}.
Global changes on scales above about $0.05 a$ appear to abide by the homogeneous strain 
assumption, hence the superposition of pair correlation functions on Fig.~\ref{fig:grphiQr}. 
Pair correlations between neighbors at smaller distances (or details of the peaks of
$g(r)$) are only approximately predicted on rescaling all distances by the same factor (as appears on Figs.~\ref{fig:zhxpPQE} and \ref{fig:rzhxpPQE}). 
And small distances of the order of $\kappa^{-1}$ (contact deflections related to forces) do not
abide by this homogeneity of strain. Otherwise, on rescaling coordinates by a factor $1-\epsilon$, 
where $\kappa^{-1}\ll \epsilon\ll 1$, one would replace any contact deflection $h$
by $\epsilon a + h$, which for $\epsilon \gg \kappa^{-1}$ would result in a much stronger 
narrowing of the force distribution than the one observed. This assumption of homogeneous strain (or affine displacements) will be further
tested on dealing with elastic moduli in paper III~\cite{iviso3}. 

The effects of a pressure reduction are more surprising. Although the evolution of solid fractions departs very little from reversibility 
(Figs.~\ref{fig:phip} and \ref{fig:phidiff}),
large initial coordination numbers in configurations A and B do not survive a pressure cycle (see Figs.~\ref{fig:zp} and \ref{fig:zpA}).
Such effects are not predicted by the
simple assumption of homogeneous expansion, which grossly fails to reproduce
 the evolution of coordination number and density on reducing the confining pressure 
(Figs.~\ref{fig:rzphipred} and \ref{fig:rzppred}). The memory of larger stresses, 
upon decompressing, imparts wider force distributions and larger friction mobilizations
in some pressure range (Fig.~\ref{fig:pdisp}), while such reductions of coordination numbers take place. 
It should be expected that decompression is less predictible,
because it is an evolution towards a larger disorder, and small differences can be amplified in the process. 
This contrasts with the compression phase, in which, for
instance, the differences between configurations A and C tend to disappear. 
Density differences are recovered on decreasing $P$, with the additional phenomenon that
new internal states at low pressure are thus being prepared, which also differ from the initially assembled ones.
While this phenomenon escapes the currently available modelling schemes,
it can be noted that  configurations with
a high coordination number, for nearly rigid grains (low pressure or high stiffness parameter $\kappa$), are extremely
rare, since each contact requires a new equation to be satisfied by the set
of sphere centre positions. Equilibrium states of rigid,
frictionless sphere assemblies, which are the initial states for configuration series A, apart from the motion of the scarce rattlers, are
isolated points in configuration space, because of isostaticity, as discussed in paper I~\cite{iviso1}. As the pressure cycle, at
the microscopic scale, is not reversible, due to friction
and to geometric changes, one should not expect such exceptional
configurations to be retrieved upon decreasing the pressure. 

We thus conclude that the internal state of granular packings, 
in addition to the assembling process, the effect of which was studied in paper I~\cite{iviso1}, 
varies according to the history of stress \emph{intensities}, even though, 
unlike in cohesive materials~\cite{WoUnKaBr05,GiRoCa07}, and in contrast with changes
in stress \emph{directions}, such loading modes only entail very small irreversible strains. 
Such commonly used characteristics of granular packings as coordination number, 
force distribution and friction mobilization level are sensitively affected by their compression history, while
strains and density changes remain very small after the assembling stage.
In particular, large
coordination numbers associated with an ideally successful suppression of friction in the sample 
preparation stage seem even more unlikely to occur generally in isotropic
sphere assemblies close to the RCP density, because 
they do not survive compression cycles. Elastic properties are studied in paper III~\cite{iviso3}, where we relate them to 
the microstucture of such states, thereby allowing for compararisons of numerical results to experimental ones. 

As possible developments of the present study, one may simulate the effects of irreversible contact deformation, 
due to material plasticity or particle breakage.

\end{document}